\documentclass[%
 aip,
 apl,
 amsmath,amssymb,
 reprint,%
]{revtex4-1}

\usepackage{graphicx}
\usepackage{dcolumn}
\usepackage{bm}

\usepackage[utf8]{inputenc}
\usepackage[T1]{fontenc}
\usepackage{mathptmx}
\usepackage{etoolbox}
\usepackage{xcolor}
\usepackage{comment}
\usepackage{tikz}
\usepackage[hidelinks]{hyperref}

\makeatletter
\def\@email#1#2{%
 \endgroup
 \patchcmd{\titleblock@produce}
  {\frontmatter@RRAPformat}
    {\frontmatter@RRAPformat{\produce@RRAP{\corrmark#1\href{mailto:#2}{#2}}}\frontmatter@RRAPformat}
  {}{}
}%

\newcommand{\corrmark}{*}
\newif\if@corrauth \@corrauthfalse
\newcommand{\corrauth}{\global\@corrauthtrue}
\gdef\@orcidid{}
\newcommand{\orcid}[1]{\gdef\@orcidid{#1}}
\newcommand{\orcidlogo}{%
 \raisebox{-0.15ex}{\tikz[baseline=-0.5ex]\node[circle,fill={rgb,255:red,166;green,206;blue,57},
 inner sep=0.5pt,text=white]{\fontsize{4.5}{4.5}\selectfont\bfseries iD};}}
\newcommand{\print@authormarks}{%
 \if@corrauth\textsuperscript{,\,\corrmark}\global\@corrauthfalse\fi
 \ifx\@orcidid\@empty\else\,\href{https://orcid.org/\@orcidid}{\orcidlogo}\gdef\@orcidid{}\fi}
\patchcmd{\doauthor}{\space \@listand}{\print@authormarks\space \@listand}{}{%
 \PackageError{main}{Could not patch \string\doauthor}{}}
\makeatother

\newcommand{\Eph}{E_{\mathrm{ph}}}
\newcommand{\Vapp}{V_{\mathrm{app}}}
\newcommand{\Vbi}{V_{\mathrm{bi}}}
\newcommand{\AlGaN}[2]{Al$_{#1}$Ga$_{#2}$N}

\begin{document}


\title[Local electric field in AlGaN $p$--$i$--$n$ photodiodes]{Predicting local electric field in Al$_{x}$Ga$_{1-x}$N $p$--$i$--$n$ photodiodes using exciton Franz--Keldysh photocurrent spectroscopy}

\author{Himel Kundu Utsha\orcid{0009-0009-4861-2587}}
\thanks{H. K. Utsha and M. Muntasir contributed equally to this work.}
\affiliation{Department of Electrical and Electronic Engineering, Bangladesh University of Engineering and Technology (BUET), Dhaka 1000, Bangladesh}

\author{Manjur Muntasir\orcid{0009-0000-4526-9634}}
\thanks{H. K. Utsha and M. Muntasir contributed equally to this work.}
\author{Md. Mohsinur Rahman Adnan\corrauth\orcid{0000-0002-3333-1514}}
 \email{mmradnan@eee.buet.ac.bd}
\affiliation{Department of Electrical and Electronic Engineering, Bangladesh University of Engineering and Technology (BUET), Dhaka 1000, Bangladesh}

\begin{abstract}
In AlGaN $p$--$i$--$n$ photodiodes, reverse bias redshifts and broadens the
sub-bandgap photocurrent responsivity through the eXciton Franz--Keldysh
(XFK) effect, and this spectral line shape encodes the local electric field
inside the device. Fitting the photocurrent spectra of two such photodiodes,
with 5\% and 40\% Al, to an XFK absorption model at each bias yields the
local voltage, $V_l$, a single parameter that sets the line shape
and uniquely determines the electric-field profile inside the device. In both
devices, $V_l$ varies linearly with the applied bias ($R^2 = 0.994$ and
$0.998$) but changes far more slowly than the bias itself, revealing a strong
local reduction of the electric field that points to non-uniform
electrostatics. The slopes show that only 12.4\% of the applied bias in the
5\% Al device, and 45.8\% in the 40\% Al device, drops across the junction at
the probed location. Extrapolated to zero bias, $V_l$ gives built-in voltages
of 3.39 and 4.29~V, within 0.02~V of the 3.41 and 4.28~V obtained from
technology computer-aided design (TCAD) simulation, which validates the
extracted local voltages at equilibrium. These results extend sub-bandgap photocurrent spectroscopy to AlGaN devices as an optical probe of local electrostatics, relevant to ultraviolet photodetectors and high-voltage electronics.
\end{abstract}

\maketitle


Al$_x$Ga$_{1-x}$N is a direct-bandgap III-nitride alloy whose bandgap can be tuned from 3.4~eV (GaN) to 6.2~eV (AlN) through the Al content $x$, placing the absorption edge anywhere from the near to the deep ultraviolet (UV).\cite{Razeghi1996,Monroy2003,Cai2021} For $x \gtrsim 0.4$ the edge falls below the 280-nm solar-blind cutoff,\cite{Cai2021} enabling intrinsically solar-blind detection without the bulky optical filters needed by Si-based detectors.\cite{Li2001,Kim2015PTL} AlGaN has thus yielded solar-blind photodiodes,\cite{Walker2000,Lambert2000} including low-noise $p$--$i$--$n$ devices with very low dark current,\cite{Li2001} and avalanche photodiodes (APDs) with high gain,\cite{Shao2014,Cai2021} exceeding $10^5$ in \AlGaN{0.05}{0.95} APDs on GaN substrates.\cite{Kim2015PTL} Because the critical electric field rises steeply with bandgap, AlGaN is also an ultrawide-bandgap platform for high-voltage electronics.\cite{Tsao2018} In both device classes, the internal field governs carrier collection, gain, dark current, and breakdown.

Yet the field inside an operating device is rarely measured. Designers instead rely on simulations that usually assume laterally uniform, one-dimensional (1D) electrostatics.\cite{Verma2020} In vertical GaN $p$--$n$ diodes, the field peaks at the outer mesa edge, where destructive breakdown usually starts, so edge terminations such as guard rings are needed to relieve field crowding.\cite{Ohta2018} In AlGaN APDs, dislocations and dry-etched sidewall features disturb the field distribution and trigger premature microplasma breakdown.\cite{Kim2015PTL,Cai2021} In solar-blind AlGaN photodiodes, two-dimensional simulations indicate possible field crowding beneath the $p$ contact, and the resistive Al-rich $p$-type layer makes the transient response depend on the illumination position.\cite{Li2001} Electroreflectance and photoreflectance spectroscopy quantify built-in fields in nitride layers and heterostructures,\cite{Wetzel1999,Janicki2018} but their oscillations decay at high fields and their analysis assumes a homogeneous field.\cite{Verma2020} Electric-field-induced second-harmonic generation maps fields in GaN devices with submicrometer resolution and shows that devices with similar terminal characteristics can have very different internal fields, with crowding wherever the edge-termination charge is uneven.\cite{Cao2021,Cao2022} Directly mapping the field inside an active device region thus remains challenging,\cite{Cao2021} and a local field probe for operating AlGaN devices is needed.

Photocurrent spectroscopy offers such a probe through the Franz--Keldysh (FK) effect. An electric field tilts the band edges, so the electron and hole wave functions overlap inside the gap and photons below the bandgap are absorbed, and the absorption edge redshifts and broadens as the field grows.\cite{Franz1958,Keldysh1958,Aspnes1966} In wide-bandgap semiconductors, the electron--hole Coulomb interaction strongly enhances this sub-bandgap absorption and changes its field dependence. This eXciton Franz--Keldysh (XFK) effect\cite{Dow1970,Blossey1970,Blossey1971,Merkulov1974} has been observed in the electroabsorption of GaN.\cite{Binet1996} Because the sub-bandgap absorption follows the local field, the bias-dependent responsivity of a reverse-biased junction encodes the field where the light is absorbed. FK photocurrent has been modeled in GaN $p$--$n$ diodes\cite{Maeda2018} and $\beta$-Ga$_2$O$_3$ Schottky diodes\cite{Maeda2024} under high reverse bias and used to extract built-in fields in an AlGaN/GaN heterostructure,\cite{Turkulets2018} and Li \textit{et al.} attributed the bias-induced redshift of an \AlGaN{0.4}{0.6} photodiode to the FK effect, assuming a uniform field.\cite{Li2001} Myers and co-workers developed the XFK effect into a quantitative local probe.\cite{Verma2020,Adnan2021,Verma2023} Verma \textit{et al.} showed that the XFK line shape of the responsivity of a GaN $p$--$n$ diode is fit with a single parameter, the local voltage $V_l$, which fixes the field profile at the illuminated position, and found a large local field reduction that they attributed to electrostatic non-uniformity.\cite{Verma2020} Adnan \textit{et al.} then identified the onset of dielectric breakdown in $\beta$-Ga$_2$O$_3$ Schottky diodes, near 6~MV/cm, from the saturation of the XFK redshift,\cite{Adnan2021} and Verma \textit{et al.} resolved the anisotropic excitonic photocurrent of $\beta$-Ga$_2$O$_3$.\cite{Verma2023}

Here, we extend XFK photocurrent spectroscopy to AlGaN and establish an XFK-based model of the internal electric-field profile of AlGaN $p$--$i$--$n$ photodiodes. Two changes are needed. First, the bandgap, exciton binding energy, Bohr radius, and permittivity in the XFK model depend on Al content. Second, practical AlGaN photodiodes are $p$--$i$--$n$ structures with an unintentionally doped (UID) absorber, so the field profile is trapezoidal rather than the triangular profile of an abrupt $p$--$n$ junction. We apply the model to the published responsivity spectra of two previously fabricated photodiodes, with 5\% and 40\% Al, validate it against technology computer-aided design (TCAD) simulation at zero bias, and show that the field at the probed location under reverse bias is far smaller than the 1D simulation predicts. To our knowledge, this is the first use of the XFK line shape to determine local fields in AlGaN devices. Because the method needs only the spectral responsivity, which is routinely measured for UV photodetectors, it can be applied across the AlGaN composition range.

\begin{figure*}
\includegraphics[width=0.80\textwidth]{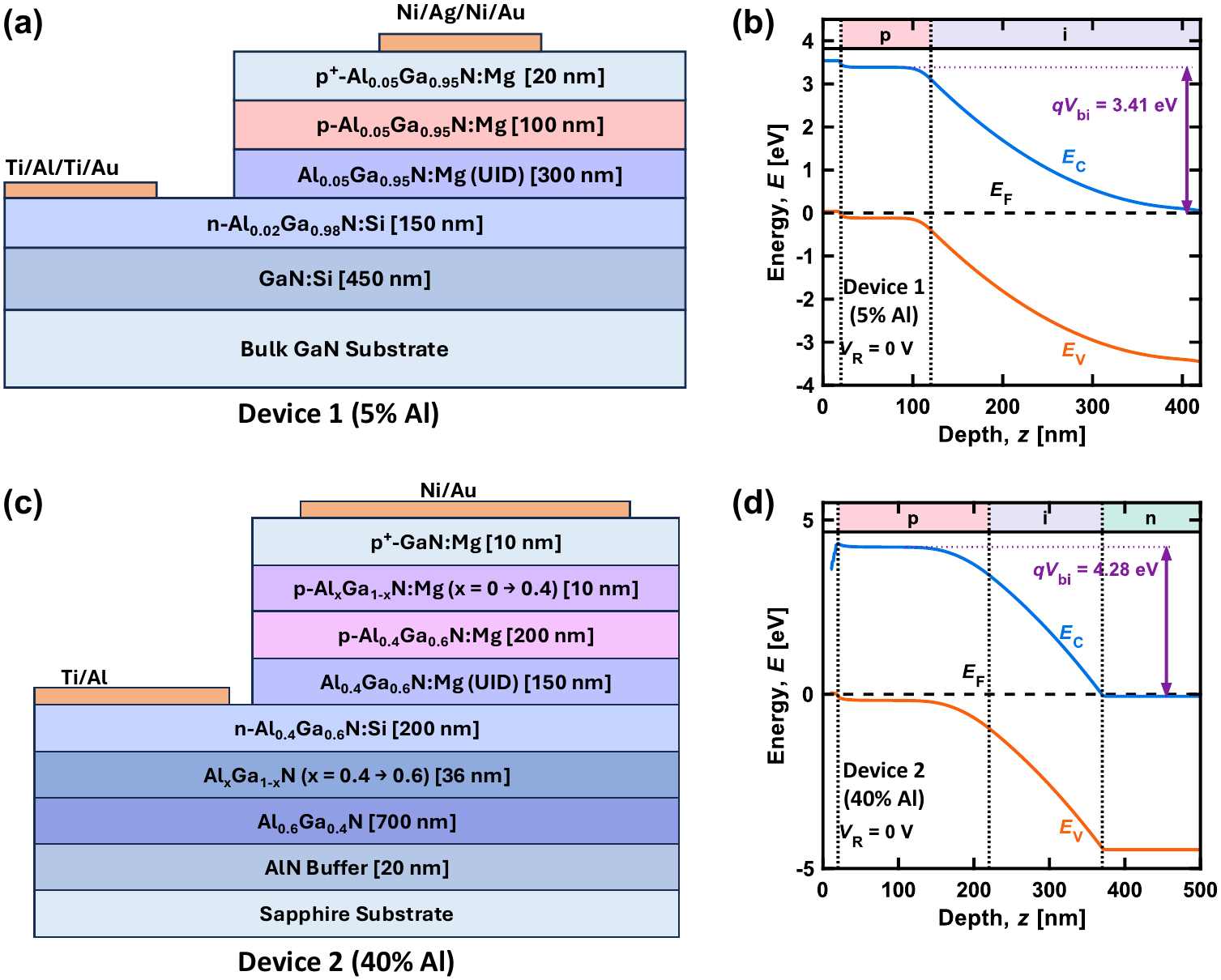}%
\caption{\label{fig:devices}Device structures and equilibrium band diagrams.
(a) Device~1: $p^{+}$-\AlGaN{0.05}{0.95}:Mg (20~nm)/$p$-\AlGaN{0.05}{0.95}:Mg
(100~nm)/UID \AlGaN{0.05}{0.95}:Mg (300~nm)/$n$-\AlGaN{0.02}{0.98}:Si
(150~nm)/GaN:Si (450~nm) on bulk GaN, with Ni/Ag/Ni/Au anode and
Ti/Al/Ti/Au cathode. (b) Simulated zero-bias band diagram of Device~1
($q\Vbi = 3.41$~eV). (c) Device~2: $p^{+}$-GaN:Mg
(10~nm)/$p$-Al$_x$Ga$_{1-x}$N:Mg graded from $x = 0$ to 0.4
(10~nm)/$p$-\AlGaN{0.4}{0.6}:Mg (200~nm)/UID \AlGaN{0.4}{0.6}:Mg
(150~nm)/$n$-\AlGaN{0.4}{0.6}:Si (200~nm)/Al$_x$Ga$_{1-x}$N graded from
$x = 0.4$ to 0.6 (36~nm)/\AlGaN{0.6}{0.4} (700~nm)/AlN (20~nm) on sapphire,
with Ni/Au anode and Ti/Al cathode. (d) Simulated zero-bias band diagram of
Device~2 ($q\Vbi = 4.28$~eV). $E_C$, $E_V$, and $E_F$ are the
conduction-band edge, valence-band edge, and Fermi level. The shaded bars
mark the $p$, $i$, and $n$ regions. }
\end{figure*}

The sub-bandgap photocurrent spectra of the two AlGaN $p$--$i$--$n$
photodiodes shown in Fig.~\ref{fig:devices} were reported by Kim \textit{et al.}\cite{Kim2015PTL} and Sood \textit{et al.}\cite{Sood2015SPIE} (Device~1) and by Li \textit{et al.}\cite{Li2001,Lambert2000} (Device~2) Here they are reanalyzed with the XFK model generalized to the $p$--$i$--$n$ field profile, from which the local voltage is obtained at each bias. Device~1 [Fig.~\ref{fig:devices}(a)]
is a top-illuminated \AlGaN{0.05}{0.95} $p$--$i$--$n$ avalanche photodiode grown by metalorganic chemical vapor deposition (MOCVD) on a free-standing $n$-GaN substrate. Its 300-nm UID absorber lies between a 100-nm $p$-\AlGaN{0.05}{0.95}:Mg layer (hole concentration
${\sim}5\times10^{17}$~cm$^{-3}$), capped by 20~nm of
$p^{+}$-\AlGaN{0.05}{0.95} ([Mg]${}\sim1\times10^{20}$~cm$^{-3}$), and a 150-nm $n$-\AlGaN{0.02}{0.98}:Si layer (${\sim}4\times10^{18}$~cm$^{-3}$). The mesas were formed by inductively coupled plasma etching, with Ni/Ag/Ni/Au anodes and Ti/Al/Ti/Au cathodes. Device~2 [Fig.~\ref{fig:devices}(c)] is a back-illuminated \AlGaN{0.4}{0.6}/\AlGaN{0.6}{0.4} solar-blind $p$--$i$--$n$ photodiode grown by low-pressure MOCVD on double-polished sapphire. Its 150-nm UID absorber
lies between a 200-nm $n^{+}$-\AlGaN{0.4}{0.6} layer and a 200-nm
$p$-\AlGaN{0.4}{0.6}:Mg layer. The $p$ layer is topped by a 10-nm compositionally graded layer of the kind used for polarization-induced hole doping\cite{Simon2010} and a 10-nm $p^{+}$-GaN contact layer, and a 700-nm
\AlGaN{0.6}{0.4} layer below the $n$ side serves as the optical window. Its mesas were defined by reactive ion etching, with Ni/Au anodes and Ti/Al cathodes. The zero-bias band diagrams [Figs.~\ref{fig:devices}(b)
and~\ref{fig:devices}(d)] were computed with Silvaco ATLAS\cite{Atlas} and give built-in voltages $\Vbi = 3.41$~V (Device~1) and 4.28~V (Device~2).

The responsivity spectra $R(\Eph)$ were acquired with a xenon lamp, a monochromator, and lock-in detection. For Device~1, the spectra are those of a 70-$\mu$m-diameter mesa under front-side illumination.\cite{Sood2015SPIE,Kim2015APEX} For Device~2, a ${\sim}25$-$\mu$m beam was focused through the sapphire substrate onto the center of the mesa.\cite{Li2001} Spectra are analyzed at reverse biases from $-10$ to $-90$~V for Device~1 and $-10$ to $-60$~V for Device~2,
in $-10$~V steps, together with the zero-bias spectrum of each device.

The $p$--$n$ photocurrent responsivity model\cite{Verma2020} is generalized
to $p$--$i$--$n$ structures:
\begin{equation}
R = \frac{q(1-r)}{\Eph}\Big[1 - a_{pn}\,e^{-\int_{-W_p}^{\,d+W_n}\alpha(F(z))\,dz}\Big],
\label{eq:R}
\end{equation}
where $q$ is the elementary charge, $r$ the surface reflectance, $\Eph$ the photon energy, $a_{pn} = e^{-(\alpha_p L_p + \alpha_n L_n)}$ the transmittance of the undepleted $p$ and $n$ portions of the collection region, of widths $L_p$ and $L_n$ and field-independent absorption coefficients $\alpha_p$ and $\alpha_n$. The layer-by-layer attenuation of the incident flux for the front-illuminated Device~1 and the back-illuminated Device~2 is given in Sec.~S1 of the supplementary material.

Here, $\alpha(F(z))$ is the field-dependent absorption coefficient of the depleted material, whose integral gives the total sub-bandgap attenuation along the optical path. The integral spans the depleted stack: $W_p$ on the $p$ side, the UID layer of thickness $d$, and $W_n$ on the $n$ side, with the depletion widths ($W_p$, $W_n$) given by Eq.~\eqref{eq:W}.

Because the band-to-band Franz--Keldysh--Aspnes model\cite{Aspnes1966} underestimates absorption within a few exciton binding energies of the gap,\cite{Dow1970,Blossey1970,Verma2020} the XFK absorption coefficient of Merkulov\cite{Merkulov1974} was used,
\begin{equation}
\begin{aligned}
\alpha(F(z)) &= \frac{C\,\xi}{\pi^{2}\left(\delta^{2}\xi^{2}+1\right)}, \qquad \delta = \Delta - 1 - \frac{9f^{2}}{2},\\
\xi &= \frac{8}{f}\exp\!\left[-\frac{4\Delta^{3/2}}{3f} - \frac{2}{\sqrt{\Delta}}\ln\!\left(\frac{8\Delta^{3/2}}{f}\right)\right],\\
f &= \frac{q|F(z)|a}{E_X}, \qquad \Delta = \frac{E_g - \Eph}{E_X},
\end{aligned}
\label{eq:XFK}
\end{equation}
where $C$ is the exciton wave-function normalization coefficient, $f$ is the field in dimensionless units, $a$ is the exciton Bohr radius, and $E_X = E_X^{0} + bF(z)^{2}$ is the Stark-shifted exciton binding energy with polarizability $b = 9q^{2}a^{2}/(8E_X^{0})$.\cite{Weiser1992} Bandgaps of $E_g = 3.50$~eV and $4.40$~eV were used for Device~1 and Device~2, respectively, taken from reported values.\cite{Sood2015SPIE,Li2001} The static dielectric constant and the electron and 
hole masses are interpolated linearly between GaN and AlN.\cite{Vurgaftman2003, Kasic2000, Barker1973, Goldberg2001} The exciton Bohr radius $a = 4\pi\varepsilon_s\hbar^{2}/(\mu q^{2})$, with $\mu$ the reduced mass,\cite{Coli2002} and the binding energy $E_X^{0} = \hbar^{2}/(2\mu a^{2})$ then follow from the same hydrogenic model, so that $a$ and $E_X^{0}$ are mutually consistent. The coefficient $C$ is found from a two-parameter fit of $C$ and $V_l$ to the highest-bias spectrum of each device and is then held fixed. Table~\ref{tab:summary} lists the values, and the supplementary material (Sec.~S3) gives the interpolation formulas.

The field enters Eq.~\eqref{eq:R} through $F(z)$. We take $z=0$ at the $p$/UID interface and $z=d$ at the UID/$n$ interface, with $d$ the UID thickness. Integrating Poisson's equation from the $p$-side depletion edge gives
\begin{equation}
F(z) = \frac{q}{\varepsilon_s}\int_{-W_p}^{\,z} N(z')\,dz',
\label{eq:F}
\end{equation}
for $-W_p \le z \le d + W_n$, where $N(z) $ is the net ionized dopant density. The depletion edges follow from charge neutrality and from the requirement that the field integrates to the local voltage,
\begin{equation}
\int_{-W_p}^{\,d+W_n} N(z)\,dz = 0, \qquad \int_{-W_p}^{\,d+W_n} F(z)\,dz = V_l.
\label{eq:constraints}
\end{equation}
For uniform acceptor density $N_A$ in the $p$ layer, net density $N_i$ in the UID layer, and donor density $N_D$ in the $n$ layer, Eqs.~\eqref{eq:F} and \eqref{eq:constraints} give
\begin{equation}
\begin{aligned}
W_p &= \frac{\sqrt{Q} - N_A d\,(N_D - N_i)}{N_A(N_A + N_D)}, \qquad W_n = \frac{N_A W_p - N_i d}{N_D},\\
Q &= N_A N_D\Big[d^{2}(N_D - N_i)(N_A + N_i)\\
&\qquad\qquad - 2\varepsilon_s (N_A + N_D)V_l/q\Big].
\end{aligned}
\label{eq:W}
\end{equation}
At the doping levels of both devices the UID layer is fully depleted at zero bias, so the field profile is trapezoidal at all biases considered and the integral in Eq.~\eqref{eq:R} spans the whole UID thickness. Equation~\eqref{eq:W} further assumes depletion edges inside the $p$ and $n$ layers. In practice we solve Eq.~\eqref{eq:constraints} numerically for the interface field, which reproduces Eq.~\eqref{eq:W} and also covers full depletion of the $p$ or $n$ layer. In that case the layer is taken as depleted across its whole thickness, with the residual field terminating in the adjacent, more heavily doped layer. The doping densities are listed in Table~\ref{tab:summary}. With the doping fixed, $V_l$ uniquely sets $W_p$, $W_n$, and $F(z)$. 

Neither the surface reflectance nor the absolute photon flux is known independently, and both vary with position on the device. Both are removed by dividing each spectrum by the zero-bias responsivity at a reference energy $E_{\mathrm{ref}}$ above the gap, where $\alpha L \gg 1$ and essentially all light entering the collection region is absorbed. The same zero-bias spectrum also fixes $a_{pn}$ at every photon energy, since the
field-dependent absorption integral is negligible at $V=0$ over the fitting range, leaving $V_l$ as the only free parameter. This gives the normalized
responsivity,\cite{Verma2020} which is derived step by step in Sec.~S2 of the supplementary material,
\begin{equation}
R^{N}(\Eph) = \frac{E_{\mathrm{ref}}}{\Eph}\Big[1 - a_{pn}\,e^{-\int\alpha(F(z))\,dz}\Big].
\label{eq:RN}
\end{equation}
Because the field-dependent absorption is taken as negligible at zero bias, the zero-bias spectrum then fixes
\begin{equation}
a_{pn}(\Eph) = 1 - \frac{\Eph}{E_{\mathrm{ref}}}\,R^{N}(\Eph; V = 0).
\label{eq:apn}
\end{equation}
We use $E_{\mathrm{ref}} = 4.40$~eV for Device~1 and 4.60~eV for Device~2.

\begin{figure*}
\includegraphics[width=0.56\textwidth]{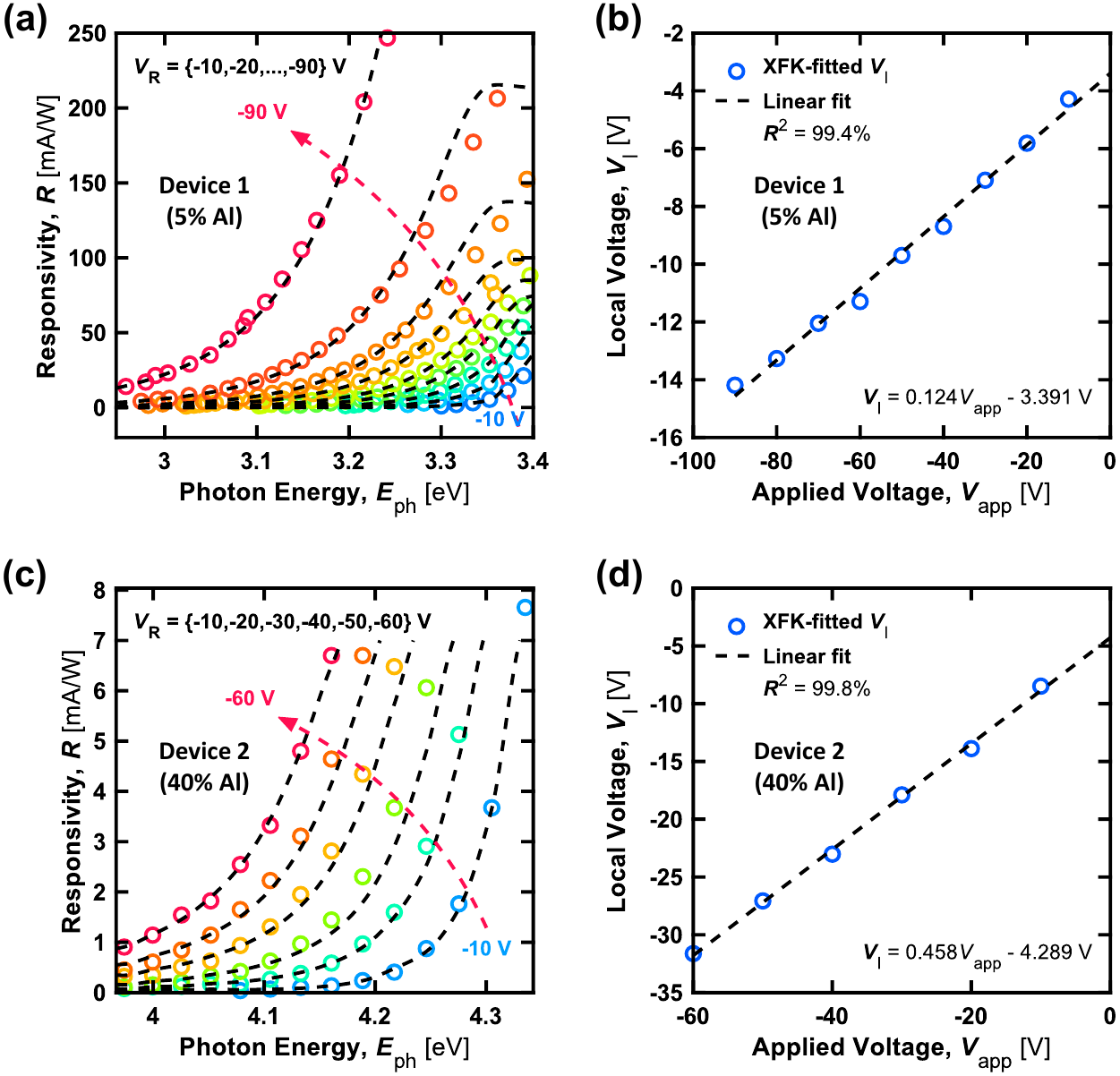}%
\caption{\label{fig:spectra}Photocurrent responsivity spectra and extracted local voltages. (a) Responsivity $R$ of Device~1 versus photon energy $\Eph$ for reverse biases $V_R = -10$ to $-90$~V in $-10$~V steps (symbols), with XFK fits (dashed lines). (b) Local voltage $V_l$ of Device~1 versus applied voltage $\Vapp$, with the linear fit $V_l = 0.124\,\Vapp - 3.391$~V ($R^2 = 0.994$) as a dashed line. (c) Same as (a) for Device~2, $V_R = -10$ to $-60$~V. (d) Same as (b) for Device~2, with $V_l = 0.458\,\Vapp - 4.289$~V ($R^2 = 0.998$). Arrows indicate increasing reverse bias. }
\end{figure*}

Figures~\ref{fig:spectra}(a) and \ref{fig:spectra}(c) show the responsivity spectra. In both devices, the sub-bandgap absorption edge shifts to lower photon energy and broadens as the reverse bias increases, the signature of field-assisted absorption. With $C$ fixed, each spectrum is fit to Eq.~\eqref{eq:RN} by minimizing the mean squared difference between the measured and modeled $\log_{10}R^{N}$, with $V_l$ as the only parameter that controls the line shape. Excellent agreement between the fitted and measured line shapes is obtained across the energy and bias ranges studied, with $R^2 \geq 0.99$.

Figures~\ref{fig:spectra}(b) and \ref{fig:spectra}(d) plot the extracted $V_l$ against the applied voltage. Both devices follow the empirical relation,\cite{Verma2020}
\begin{equation}
V_l = c_l\,\Vapp - \Vbi,
\label{eq:linear}
\end{equation}
where $c_l$ is the fraction of the applied bias that appears across the junction at the probed location. Linear fits give $c_l = 0.124$ and $\Vbi = 3.391$~V for Device~1 ($R^2 = 0.994$), and $c_l = 0.458$ and $\Vbi = 4.289$~V for Device~2 ($R^2 = 0.998$). The extrapolated built-in voltages agree with the TCAD values of 3.41 and 4.28~V to within 0.019 and 0.009~V, respectively.

\begin{table}
\caption{\label{tab:summary}Model inputs and extracted parameters.}
\begin{ruledtabular}
\begin{tabular}{lcc}
 & Device 1 & Device 2 \\
\hline
Al fraction ($p$ and UID layers) & 0.05 & 0.40 \\
UID thickness $d$ (nm) & 300 & 150 \\
$N_A$, $p$ layer (cm$^{-3}$) & $5\times10^{17}$ & $1\times10^{17}$ \\
$N_i$, UID layer (cm$^{-3}$) & $+4\times10^{16}$ & $-3.5\times10^{16}$ \\
$N_D$, $n$ layer (cm$^{-3}$) & $4\times10^{18}$ & $1\times10^{19}$ \\
$E_g$ (eV) & 3.50 & 4.40 \\
$\varepsilon_s/\varepsilon_0$ & 10.31 & 9.64 \\
$\mu/m_0$ & 0.18 & 0.22 \\
$a$ (nm) & 3.01 & 2.28 \\
$E_X^{0}$ (meV) & 23.22 & 32.71 \\
$C$ & $4.4\times10^{8}$ & $5\times10^{8}$ \\
\hline
$\Vbi$, TCAD (V) & 3.41 & 4.28 \\
$\Vbi$, XFK intercept (V) & 3.391 & 4.289 \\
Bias fraction $c_l$ & 0.124 & 0.458 \\
$R^2$ of Eq.~\eqref{eq:linear} & 0.994 & 0.998 \\
Peak $|F|$, XFK/TCAD (MV/cm)\footnote{At $V_R = -90$~V (Device 1) and $-60$~V (Device 2).} & 0.52/2.43 & 1.08/2.00 \\
\end{tabular}
\end{ruledtabular}
\end{table}

\begin{figure*}
\includegraphics[width=0.94\textwidth]{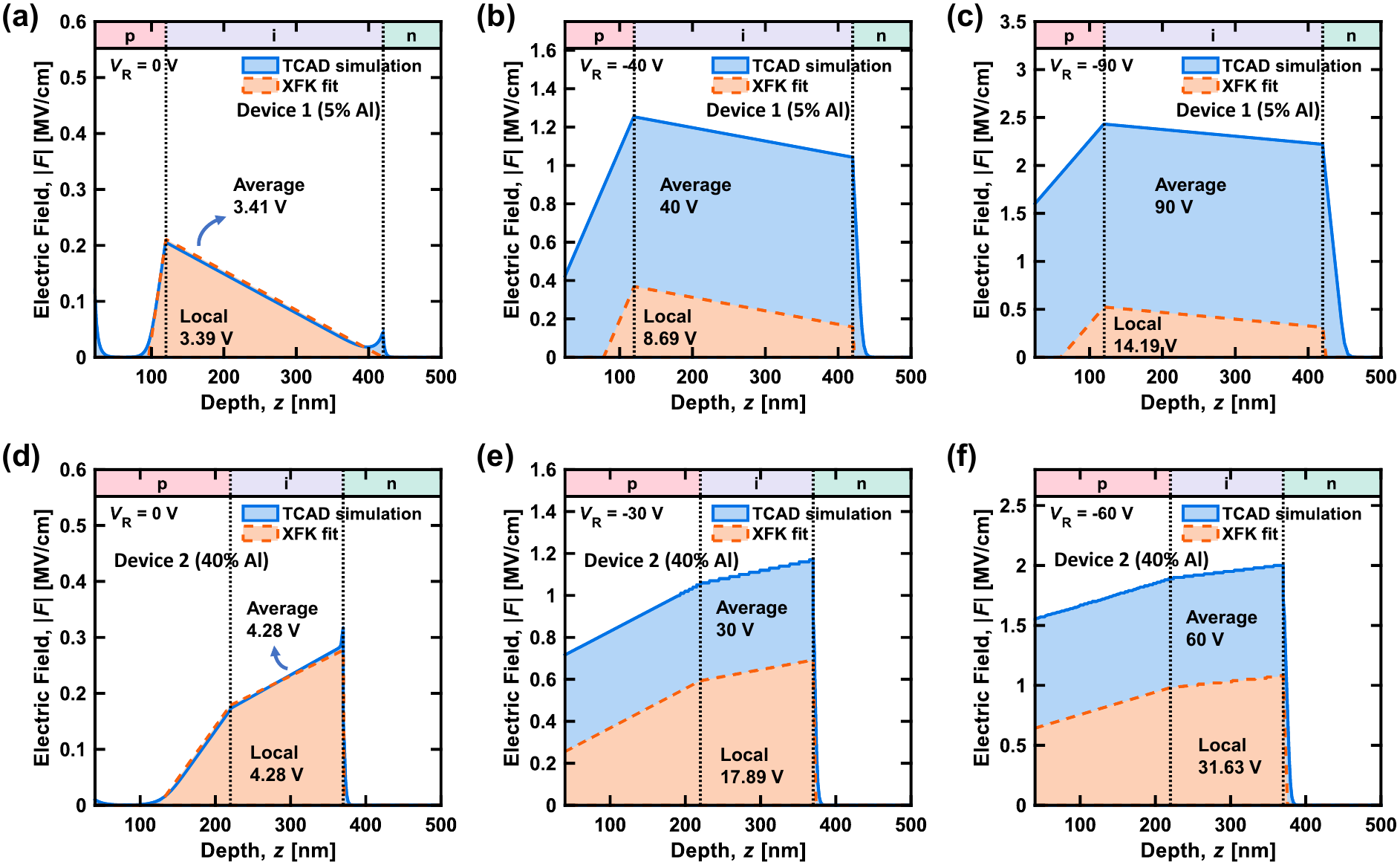}%
\caption{\label{fig:field}Vertical electric-field profiles $|F(z)|$ at the probed location. The blue curves are 1D TCAD simulations and the orange curves are profiles reconstructed from the XFK fit through Eqs.~\eqref{eq:F}--\eqref{eq:W}. Panels (a)--(c) show Device~1 at $V_R = 0$, $-40$, and $-90$~V, and panels (d)--(f) show Device~2 at $V_R = 0$, $-30$, and $-60$~V. The TCAD labels give $|V_R|$ (and $\Vbi$ at $V_R = 0$). The XFK labels give the fitted total local voltage $|V_l|$, extrapolated with Eq.~\eqref{eq:linear} at $V_R = 0$. Dotted lines mark the $p$/UID and UID/$n$ interfaces.}
\end{figure*}

Figure~\ref{fig:field} compares the field profiles reconstructed from the fitted $V_l$ with the 1D TCAD profiles. At zero bias the two coincide in both devices [Figs.~\ref{fig:field}(a) and \ref{fig:field}(d)]. The local voltages extrapolated to zero bias with Eq.~\eqref{eq:linear}, $V_l = -3.39$ and $-4.29$~V, equal $-\Vbi$, and the peak fields of about 0.2 and 0.28~MV/cm match TCAD. This agreement in equilibrium, where no external current flows, is consistent with the material parameters used in Eq.~\eqref{eq:XFK} and the electrostatic model of Eqs.~\eqref{eq:F}--\eqref{eq:W}. The trapezoidal shape of the profiles reflects the residual charge in the UID layer. The field decreases toward the $n$ side in Device~1 and increases toward it in Device~2, corresponding to net positive and net negative space charge, respectively, in the electrostatic model. In Device~2 the low ionized acceptor density of $p$-\AlGaN{0.4}{0.6}, a consequence of the deep Mg acceptor level in Al-rich AlGaN,\cite{Nam2003} lets the depletion region extend deep into the $p$ layer.

Under reverse bias the profiles diverge. For Device~1, the local peak field is about 0.36~MV/cm at $-40$~V and 0.52~MV/cm at $-90$~V, compared with 1.25 and 2.43~MV/cm from TCAD [Figs.~\ref{fig:field}(b) and \ref{fig:field}(c)]. For Device~2, it is about 0.69 and 1.08~MV/cm at $-30$ and $-60$~V, compared with 1.17 and 2.00~MV/cm [Figs.~\ref{fig:field}(e) and \ref{fig:field}(f)]. The 1D simulation therefore overestimates the peak field at the probed location by factors of about 4.7 and 1.9 at the highest biases. The local peak field remains well below the TCAD peak field at every bias (Sec.~S4 of the supplementary material). A similar reduction, corresponding to $c_l \approx 0.23$, can be inferred from the local voltage and built-in voltage reported for a GaN $p$--$n$ diode,\cite{Verma2020} where it was attributed to electrostatic non-uniformity associated with the electrode geometry.

Several mechanisms can make $c_l < 1$ at a given location. Field fringing near the contact and mesa edges redistributes the applied bias. Lateral current flow through the resistive $p$-type AlGaN can drop part of the bias between the contact and the illuminated region. Li \textit{et al.} attributed the strong position dependence of the pulse response of Device~2 to the high resistivity of its $p$-\AlGaN{0.4}{0.6} layer.\cite{Li2001} Charge trapped at the surface or mesa sidewall can also screen part of the applied field. Measuring $V_l$ as a function of distance from the anode edge, for several contact geometries and temperatures, would separate these contributions. Whatever the mechanism, the result has practical consequences. In photodetectors, the XFK tail sets a bias-dependent long-wavelength cutoff, so the field actually present in the absorber determines the out-of-band rejection. In high-voltage devices, local fields that differ several-fold from 1D predictions mean that breakdown hot spots cannot be located from simulation alone.

In summary, we extended exciton Franz--Keldysh photocurrent spectroscopy from GaN $p$--$n$ diodes to two previously reported AlGaN $p$--$i$--$n$ photodiodes with 5\% and 40\% Al. A closed-form $p$--$i$--$n$ electrostatic model lets each responsivity spectrum be fit with the local voltage as the only parameter that controls the line shape. The extracted local voltage is linear in bias, its zero-bias intercept reproduces the built-in voltage within 0.02~V, and at zero bias the reconstructed field matches TCAD. Under reverse bias, only 12.4\% and 45.8\% of the applied bias acts at the probed location, and at the highest biases the local peak field is about 2--5 times below the 1D estimate. Sub-bandgap photocurrent spectroscopy therefore offers a contactless way to map electric fields in AlGaN devices and to test electrostatic designs.


\section*{Supplementary Material}
See the supplementary material for the layer-by-layer photon-flux chains of both devices, the step-by-step normalization of the photocurrent spectra, the exciton binding energy and Bohr radius as functions of Al content, and the relation between the extracted local peak field and the TCAD peak field.

\begin{acknowledgments}
The authors gratefully acknowledge Bangladesh University of Engineering and Technology (BUET) for supporting this work.
\end{acknowledgments}

\section*{Author Declarations}
\subsection*{Conflict of Interest}
The authors have no conflicts to disclose.

\subsection*{Author Contributions}
\textbf{Himel Kundu Utsha}: Formal analysis (equal); Methodology (equal); Software (equal); Visualization (equal); Writing -- original draft (equal); Writing -- review \& editing (equal). \textbf{Manjur Muntasir}: Formal analysis (equal); Methodology (equal); Software (equal); Visualization (equal); Writing -- original draft (equal); Writing -- review \& editing (equal). \textbf{Md. Mohsinur Rahman Adnan}: Conceptualization (lead); Supervision (lead); Writing -- review \& editing (equal).

\section*{Data Availability}
The data that support the findings of this study are available from the corresponding author upon reasonable request.

\bibliography{references}

\end{document}